\documentclass[prl,preprint,tightenlines,nofootinbib]{revtex4}
 
\begin{document}
 
 \title{Mass Dependence of Vacuum Energy}
 
\author{S. A. Fulling}
\email{fulling@math.tamu.edu}
 \homepage{http://www.math.tamu.edu/~fulling}
\affiliation{Departments of Mathematics and Physics, Texas A\&M 
University, 
  College Station, TX, 77843-3368 USA}

 \date{July 21, 2005}

\begin{abstract}
The regularized vacuum energy (or energy density) of a quantum 
field subjected to static external conditions is shown to satisfy a 
certain partial differential equation with respect to two 
variables, the mass and the ``time'' (ultraviolet cutoff 
parameter).
 The equation is solved to provide  integral expressions for the 
regularized energy (more precisely, the cylinder kernel)
  at positive mass in terms of that for zero mass. 
 Alternatively, for fixed positive mass 
 all coefficients in the short-time asymptotics of 
the regularized energy  can be obtained 
recursively from the first nontrivial coefficient, which is the 
 renormalized vacuum energy.
  \end{abstract}      
 \maketitle

\section{Background}

 The experimentally documented physical relevance of vacuum energy 
involves the electromagnetic field (the Casimir effect).
 Other theoretical work also usually considers massless fields
 (scalar fields for calculational and conceptual simplicity;
 neutrinos or gluons for more exotic or speculative applications).
 Nevertheless, the vacuum energy of a quantum field with mass is of 
theoretical importance.
 Adding the mass term is the simplest generalization of the basic 
model of a massless scalar field in an empty flat cavity,
 a step toward the more complicated scenarios with space-dependent 
 external potentials or curved background geometries.
Although massive Casimir-type effects in the simplest geometries 
 tend to be exponentially 
damped relative to their massless counterparts, they can be less
negligible in other geometries \cite[Secs.\ 7.4--7.5]{K}.
 
 The massive field in one-dimensional space was studied by 
 Hays \cite{Hays}.
 (See also [\onlinecite{BH,Kay,AW}].)
 Perhaps the most notable feature of the results is the presence of 
a logarithmic divergence, absent from the massless case.
 Although physically harmless in the context of the one-dimensional 
bag model \cite{BH,Hays}, a logarithmic divergence is regarded as 
particularly problematic by some theorists because it is not 
automatically eliminated by dimensional or zeta-function 
regularization.

 As I have stressed elsewhere \cite{FG,systemat,norman},
 vacuum energy is one of a series of moments of the spectral 
distribution of the differential operator appearing in the field 
equation; 
  these quantities arise as the 
coefficients in the short-time asymptotics of a certain Green 
function, the ``cylinder kernel''\negthinspace.
 The diagonal value  of the latter is,
 up to a differentiation and a numerical factor, the regularized 
vacuum energy defined by an ultraviolet cutoff (with the time 
variable as the cutoff parameter) \cite{BH,Hays,CVZ}.
 Thus the vacuum energy (regardless of experimental relevance) is 
of mathematical interest as a tool of spectral analysis.
 The cylinder-kernel coefficients incorporate nonlocal geometrical 
information that is not extractible from the much-studied 
 short-time asymptotic expansion of the heat kernel.
(The cylinder kernel is minus twice the $t$-derivative of the 
Green function 
employed in [\onlinecite{Hays,BH}].)

 \section{Notation}

 Consider a field equation of the type 
 $\frac{\partial^2\phi}{\partial t^2} = -H\phi$, where
 \begin{equation}
 H=H_0 + \mu, \quad \mu\equiv m^2,
\label{operator} \end{equation}
 and $H_0$ is a self-adjoint second-order differential operator in 
the spatial variables, such as $-\nabla^2$.
 For simplicity (although the treatment of local energy density is 
actually more general) assume that the spatial domain is compact 
and $H$ has a discrete, positive spectrum $\{\omega_n^2\}$ with 
orthonormal eigenfunctions $\{\phi_n(x)\}$.

  For auxiliary mathematical purposes one studies the 
 \emph{heat kernel}
 \begin{equation}
 K(t,x,y) = \langle x|e^{-tH}|y\rangle
 = \sum_{n=1}^\infty e^{-t\omega_n^2} \phi_n(x) \phi_n(y)^*
 \label{heat}\end{equation}
 and the \emph{cylinder (Poisson) kernel}
 \begin{equation}
 T(t,x,y) = \langle x|e^{-t\sqrt{H}}|y\rangle
 = \sum_{n=1}^\infty e^{-t\omega_n} \phi_n(x) \phi_n(y)^* .
 \label{cyl}\end{equation}
Each of these can be ``traced'' over space; for example,
 \begin{equation}
 \mathop{\rm Tr} T 
 = \int \langle x|e^{-t\sqrt{H}}|x\rangle \, dx
 = \sum_{n=1}^\infty e^{-t\omega_n}.
\label{cyltrace}\end{equation}
 Formally,
 the vacuum energy of the quantized field configuration is
 \begin{equation}
 E = \frac12 \sum_{n=1}^\infty \omega_n
 = -\, \frac12 \lim_{t\to 0} \frac{\partial}{\partial t}
  \mathop{\rm Tr} T,
 \label{energy}\end{equation}
 and one possible definition (see [\onlinecite{systemat,norman}])
 of the vacuum energy density is
\begin{equation}
 T_{00}(x) 
 = \frac12 \sum_{n=1}^\infty \omega_n \phi_n(x) \phi_n(y)^*
 =-\,\frac12 \lim_{t\to 0}\frac{\partial}{\partial t} T(t,x,x).
 \label{density}\end{equation}

  In reality, the limits in (\ref{energy}) and (\ref{density}) do 
not exist,  but    
  $\,\mathop{\rm Tr} T$ and $T(t,x,x)$ possess asymptotic 
expansions as $t\downarrow 0$ of the form \cite{CVZ,GG,FG,BM}
\begin{equation}  T \sim
\sum_{s=0}^\infty e_s t^{-d+s}
+\sum^\infty_{\scriptstyle s=d+1\atop
\scriptstyle s-d \mathrm{\ odd}} f_s t^{-d+s} \ln t,
 \label{cylseries}\end{equation}
where the coefficients of the divergent terms are simple, local 
objects that can be absorbed by renormalization.
 (Here $d$ is the spatial dimension.)
 Therefore, one regards (\ref{cyl}) and  (\ref{cyltrace}),
 after operation by $-\frac12 \frac{\partial}{\partial t}\,$,
  as the \emph{regularized} energy and energy density, and
 one regards
 $-\frac12$ times the coefficient of the term of order $t$ in 
(\ref{cyl}) and  (\ref{cyltrace}) as the 
 \emph{renormalized} energy and energy density:
 \begin{equation}
 E   \mbox{ or } T_{00} = - {\textstyle \frac12} e_{d+1}.
 \label{renorm}\end{equation}

 Similarly, if $K$ stands for either  
 $\,\mathop{\rm Tr} K$ or $K(t,x,x)$,
 it has an expansion of the form
\begin{equation}
 K \sim \sum_{s=0}^\infty b_s t^{(-d+s)/2}.
 \label{heatseries}\end{equation}

  \section{The main equation} 

The coefficients in (\ref{cylseries}) and (\ref{heatseries}) are
 functions of $\mu$.
 Let us write $T(\mu,t)$ and $K(\mu,t)$ for the quantities being 
expanded and write $e_s(\mu)$, etc., for the coefficients.
 In the case of the heat kernel, it is elementary that
 \begin{equation}
 K(\mu,t)= K(0,t) e^{-\mu t},
 \label{heatfactor}\end{equation}
and
 from (\ref{heatfactor}) it is routine to find formulas for the 
$b_s(\mu)$ in terms of $b_{s'}(0)$ ($s'\le s$).
 For the cylinder kernel it is clear that no elementary 
factorization like (\ref{heatfactor}) occurs, 
 and hence the mass dependence is much more interesting and nontrivial.

On the other hand, (\ref{heatfactor}) is equivalent to the 
differential equation
 \begin{equation}
 \frac{\partial K}{\partial \mu} = -tK.
 \label{heatpde}\end{equation}
 The goal of the present paper is to find, as nearly as possible, 
an analogue of (\ref{heatpde}) for the 
 quantities $T(\mu,t)$ related to vacuum energy. 
 Since $\omega_n(\mu) = \sqrt{\omega_n(0)^2+\mu}$,
 it is easy to show from (\ref{cyltrace}) or (\ref{cyl}) that
 \begin{equation}
\frac{\partial^2}{\partial\mu\,\partial t} \left(\frac Tt \right)
 =\frac T2 \,,
 \label{cylpde}\end{equation}
which is the central equation of this paper.
  The variables $t$ and $\mu$ naturally range from $0$ to 
  $+\infty$.

 If its right side  were zero, (\ref{cylpde}) would 
be mathematically equivalent to the massless wave equation in 
 two-dimensional space-time written in null (light cone) 
coordinates; as is well known, its general solution would then be
 $T(\mu,t)/t=A(t) + B(\mu)$, where $A$ and $B$ are arbitrary 
functions. 
The full equation (\ref{cylpde}) is of the same hyperbolic type, 
and one can again expect the general solution to involve two 
arbitrary one-variable functions.
 One of these should be the ``initial value'' $T(0,t)$,  in analogy 
with  (\ref{heatfactor}).
 The remaining boundary condition is 
 (cf.\ (\ref{cyl})--(\ref{cyltrace}))
 \begin{equation}
 \lim_{t\to+\infty} T(\mu,t) =0.
 \label{zeroinfty}\end{equation}

 \section{Solution by Laplace transform}

 Let $F(s,t)$ be the Laplace transform of $T(\mu,t)/t$ with respect 
to~$\mu$.
 Then (\ref{cylpde}) is equivalent to
 \[
 s\,\frac {dF}{dt} - \frac{\partial}{\partial t}\,
  {T(0,t)\over t}= \frac t2\, F\,;
 \]
 i.e.,
 \begin{equation}
\frac {dF}{dt} - \frac t{2s}\, F = \frac{\partial}{\partial t}\, 
 {T(0,t)\over st}\,.
 \label{ltpde}\end{equation}
 The solution of (\ref{ltpde}) consistent with
  (\ref{zeroinfty}) is 
 \begin{equation}
  F(s,t) = -e^{t^2/4s}\int_t^\infty e^{-v^2/4s}
\frac{\partial}{\partial v}\, {T(0,v)\over sv}   \,dv. 
 \label{lapcyl}\end{equation}
 Equivalently,
  \begin{equation}
  F(s,t) = {T(0,t)\over st}
 -\frac1{2s^2} e^{t^2/4s}\int_t^\infty e^{-v^2/4s}
T(0,v)\,dv. 
 \label{lapcylbp}\end{equation}
Thus, in principle, $T(\mu,t)$ can be calculated from $T(0,v)$.

 Indeed, the inverse Laplace transform can be performed at the 
kernel level (under the integral sign in (\ref{lapcyl}) or
(\ref{lapcylbp})) \cite[p.\ 1026]{AS}:
 \begin{equation} 
 {T(\mu,t)\over t} =
 -\int_t^\infty J_0\bigl(m\sqrt{v^2-t^2}\bigr) 
\frac{\partial}{\partial v}\left({T(0,v)\over v}\right) dv
 \label{bescyl} \end{equation}
 or
 \begin{equation}
 T(\mu,t)= T(0,t)
  -t\int_t^\infty \frac {m\,dv}{\sqrt{v^2-t^2}} \, 
 J_1\bigl(m\sqrt{v^2-t^2}\bigr) T(0,v).
 \label{bescylbp}\end{equation}
 A change of variable
 ($w^2=v^2-t^2$)
  converts (\ref{bescylbp}) to
 \begin{equation}
 T(\mu,t)= T(0,t)
 -t\int_0^\infty \frac {m\,dw}{\sqrt{w^2+t^2}} \, J_1(mw) 
T\bigl(0,\sqrt{w^2+t^2}\bigr),
 \label{bescylvar}\end{equation}
 and there is a similar variant of (\ref{bescyl}).

 \subsection{Example 1}  The cylinder kernel of the free massless 
scalar field in spatial dimension~$d$ is 
 \begin{equation}
T(0,t,\mathbf x,\mathbf y)= {\Gamma(c)\pi^{-c} t\over (t^2+z^2)^c}
 \label{freemassless}\end{equation}
 where $z\equiv |\mathbf x-\mathbf y|$ is the spatial separation
 and $c = \frac12(d+1)$.
 According to (\ref{bescyl}), therefore,
 \begin{equation}
T(\mu,t,\mathbf x,\mathbf y)=2c\Gamma(c)\pi^{-c} t \int_0^\infty
 {w J_0(mw) \,dw\over (w^2 + t^2 + z^2)^{c+1}}\,.
\label{freebes}\end{equation}
 From \cite[p.~425]{W} follows
 \begin{equation}
T(m^2,t,\mathbf x,\mathbf y)=   2^{1-c} \pi^{-c} m^c t (t^2+z^2)^{-c/2}
K_c(m\sqrt{t^2+z^2}),
 \label{free}\end{equation}
 which is the correct formula\footnote{The Bender--Hays Green 
function \cite{Hays,BH}
is the Green function of the Helmholtz equation in one higher 
dimension,
 $[-\nabla^2 -\frac{\partial^2}{\partial t^2} +m^2]G = 
\delta(t)\delta^d(\mathbf{z})$.
 In the free case this kernel is known \cite[(4.25)]{OCF} and is 
proportional to $K_{c-1}(m\sqrt{t^2+z^2})$.
 Differentiation and a recursion relation for the modified Bessel 
function then lead to (\ref{free}).}
 for the cylinder kernel of the free field of mass $m$. 
 All the solution formulas for problems with infinite flat boundaries
 now follow by the method of images.

 \subsection{Example 2} Let $d=1$ and consider an interval of 
length~$L$ with Dirichlet boundary conditions.
 For the massless case the solution by images can be summed in 
closed form \cite[(23) and (27)]{FG} with the result
 \begin{equation}
  \mathop{\rm Tr} T(0,t) = \frac12 \,{\sinh (\pi t/L) \over
 \cosh(\pi t/L) -1} -\frac12\,.
 \label{cyltr1}\end{equation}
 We apply (\ref{bescylvar}) and (\ref{energy})
  and compare with the conclusions 
of Hays \cite{Hays} about the renormalized total energy in the 
massive case
 (finding complete agreement).
 It is convenient to separate out the contribution of the free Green 
function (\ref{freemassless})  (with $c=1$, $z=0$, and
 integrated over $0<x<L$),
   because that is where all the divergences lie.
 That is, in both places in (\ref{bescylvar}) write
 \[
 T(0,v) = {L\over\pi v} +\left[T(0,v) - {L\over\pi v}\right].
 \]
 When we apply the operator
 $-\frac12 \lim_{t\to0}\frac{\partial}{\partial t}$ to 
(\ref{bescylvar})
we thus encounter four terms:
\begin{itemize}

\item The divergent term 
 \begin{equation}
\frac L{2\pi t^2}
 \label{masslessren}\end{equation}
 (present already in empty 
space) is the mass-independent 
part of the renormalization of the bag constant in 
[\onlinecite[(3.10)]{Hays}].

\item The remaining (bracket) contribution of the first term in 
(\ref{bescylvar}) is the familiar massless Casimir energy,
 \begin{equation}
-\,\frac{\pi}{24L}\,. 
 \label{massless}\end{equation}
  (It comes from the $O(t^2)$ term in the Taylor 
expansion of (\ref{cyltr1}).  Regrettably, that crucial term is 
written in [\onlinecite[(27)]{FG}] with the wrong sign.) 

\item The contribution of the free Green function to the integral 
in (\ref{bescylvar}) works out to
\[
{m^2Lt\over 2\pi}\,\ln\left({mt\over2}\right) 
+(2C-1){m^2Lt\over 4\pi}
\]
($C= 0.577\ldots\,$, Euler's constant).
The corresponding term in the regularized energy,
\begin{equation} 
-\, {m^2L\over 4\pi}\,\ln\left({mt\over2}\right) 
-(2C+1){m^2L\over 8\pi} \,,
\label{massren}\end{equation} 
  is the 
mass-dependent part of the renormalization  
[\onlinecite[(3.10)]{Hays}].
It, also, is present in empty space.
(It includes a finite term, proportional to $m^2L$, which is 
actually ambiguous in the sense that the scale factor in the 
argument of the logarithm function is arbitrary.)

\item The remaining (bracket)  part of the integral splits into 
two disparate pieces.

\begin{itemize}
\item The term $-\frac12$ in (\ref{cyltr1}) 
contributes\footnote{This follows from [\onlinecite[(6.552.1)]{GR}]
 and the power series of the Bessel functions.}
\begin{equation}
 -\,\frac14 \lim_{t\to0} \frac{\partial}{\partial t}(1-e^{-mt})
 =-\,\frac m4
\label{massconst}\end{equation}
to the energy, in agreement with [\onlinecite[(3.15)]{Hays}].
This constant term, associated with paths that reflect from the 
boundaries an odd number of times \cite[Sec.~4]{Hays}, represents 
the energy of interaction of the massive field with the two 
boundaries separately (i.e, it survives when $L$ approaches 
infinity, and it does not contribute to the Casimir force).

\item What remains is the contribution of the paths that reflect 
an even number of times; it is the mass-dependent part of the 
true Casimir energy.
In our present approach it equals\footnote{The 
error made by setting $t=0$ inside the 
integrand of this term 
of (\ref{bescylvar}) before differentiating is of order $t^2\ln t$, 
so it vanishes in the limit.}
\[
\frac m4 \int_0^\infty {J_1(mw)\over w}\,
\left[ {\sinh(\pi w/L)\over \cosh(\pi w/L)-1} -
{2L\over \pi w}\right] dw 
 \]
 \begin{equation}
{} =\frac m4 \int_0^\infty J_1\left({mLu\over \pi}\right)
 \left[\coth \left(\frac{u}2\right) -\frac2u \right] {du\over u}
\,.
\label{energyint}\end{equation}
In Hays's approach [\onlinecite[(3.11) and (3.13)]{Hays}] it 
appears as
\begin{equation}
-\,\frac m{2\pi} \sum_{s=1}^\infty {K_1(2mLs)\over s} + {\pi\over 
24L}
\label{energysum}\end{equation}
(since Hays's sum includes the massless Casimir energy, 
(\ref{massless})).
 The equivalence of (\ref{energyint}) and (\ref{energysum}) is not 
obvious, but it can be verified by the method of
 \cite[p.~427]{W}.
(It has also been tested numerically.)
   We claim no practical advantage for the integral, since 
the sum converges faster.
\end{itemize}
\end{itemize}

 \section{Partial solution by  recursion}

  Rarely will one of the integrals 
 (\ref{lapcyl})--(\ref{bescylvar})  be evaluatable analytically 
in any particular case.
 Moreover, $T(0,t)$ often will not be available for 
arbitrarily large~$t$. 
It is worthwhile, therefore, to see how much information can be
 obtained from the known asymptotic structure (\ref{cylseries})
 if the coefficients for $m=0$ are known.

 By substituting (\ref{cylseries}) into (\ref{cylpde}) one obtains
 \begin{widetext}
 \[
 \sum_{s=0}^\infty (-d+s-1)\frac{\partial e_{s}}{\partial\mu} t^{-
d+s-2}
+\sum^\infty_{\scriptstyle s=d+1\atop
\scriptstyle s-d \mathrm{\ odd}} (-d+s-1) \frac{\partial f_{s}}{\partial\mu} 
 t^{-d+s-2} \ln t
+\sum^\infty_{\scriptstyle s=d+1\atop
\scriptstyle s-d \mathrm{\ odd}}  \frac{\partial f_{s}}{\partial\mu} 
 t^{-d+s-2} 
 \]
\begin{equation} {} =
\sum_{s=2}^\infty \frac{e_{s-2}}2 t^{-d+s-2}
+\sum^\infty_{\scriptstyle s=d+3\atop
\scriptstyle s-d \mathrm{\ odd}} \frac{f_{s-2}}2 t^{-d+s-2} \ln t.
 \label{cylrr}\end{equation}
 \end{widetext}
 Therefore, we have the recursion relations
 \begin{equation}
(-d+s-1)\frac{\partial f_{s}}{\partial\mu} = \frac{f_{s-2}}2
 \label{rrf}\end{equation}
 for $s  -d$ odd and positive, and
 \begin{equation}
(-d+s-1)\frac{\partial e_{s}}{\partial\mu}
  =\frac{e_{s-2}}2 -\frac{\partial f_{s}}{\partial\mu}
 \label{rre}\end{equation}
 for all nonnegative integers $s  $, 
 the  terms being set to $0$ when not defined.
 Generically these equations can be solved recursively for 
$f_{s  }(\mu)$ and $e_{s  }(\mu)\,$, respectively
 (the initial data $f_{s  }(0)$ and $e_{s  }(0)$ being presumed 
known).
Exceptions occur when $-d+s-1=0$ (i.e., $s=d+1$);
   then  (\ref{rrf}) becomes a tautology and
 (\ref{rre}),
\begin{equation}
\frac{\partial f_{d+1}}{\partial\mu} = \frac{e_{d-1}}2 \,,
\label{excep}\end{equation}
takes its place as the equation determining $f_{d+1}$.

 Thus there is no equation to determine $e_{d+1}(\mu)$.
 The reason is that there is no way in this approach to 
impose the second boundary condition (\ref{zeroinfty}), so 
the solution must involve an arbitrary function.
 Ironically, that function turns out to be naturally 
identified with the renormalized vacuum energy 
(\ref{renorm}), precisely the quantity of greatest 
physical interest.
 In fact, the only coefficients that have been completely 
determined by this exercise are the ones that are 
equivalent to heat-kernel coefficients 
\cite{FG,systemat,norman}.
 Nevertheless, the calculation clarifies the structure of 
the problem and shows that once $e_{d+1}(\mu)$ is known
 (along with the mass-zero coefficients), all the higher 
cylinder-kernel coefficients are computable.

\bigskip
 \begin{acknowledgments}
 I thank Stuart Dowker and Klaus Kirsten for comments on the first 
draft, and Todd Zapata for help with the numerical verification 
of ``(\ref{energyint}) = (\ref{energysum})''\negthinspace.
 \end{acknowledgments}

\goodbreak


\begin{thebibliography}{00}
\frenchspacing

 \bibitem{K} K. Kirsten, 
 \textit{Spectral Functions in Mathematics and Physics}
 (Chapman \& Hall/CRC, Boca Raton, 2002).

 \bibitem{Hays} P. Hays, Ann. Phys. (NY) {\bf 121}, 32 (1979).

 \bibitem{BH} C. M. Bender and P. Hays, Phys. Rev. D {\bf14}, 2622 
(1976).

 \bibitem{Kay} B. S. Kay, Phys. Rev. D {\bf20}, 3052 (1979).

 \bibitem{AW} J. Ambj{\o}rn and S. Wolfram,
Ann. Phys. (NY) {\bf 147}, 1 (1983). 

   \bibitem{FG}   S. A. Fulling and R. A. Gustafson,
 Electronic J. Diff. Eqs. {\bf1999}, No. 6 (1999).

  \bibitem{systemat} S. A. Fulling, J. Phys. A {\bf36},  6857 (2003).

 \bibitem{norman}   S. A. Fulling, in
\textit{Quantum Field Theory Under the Influence of External Conditions},
ed. by  K. A. Milton (Rinton Press, Princeton, 2004),
pp. 166--174.


 \bibitem{CVZ}  G. Cognola, L. Vanzo, and S. Zerbini,
  J. Math. Phys. {\bf33}, 222 (1992).

 \bibitem{GG} P. B. Gilkey and G. Grubb,
 Commun. Partial Diff. Eqs. {\bf23}, 777 (1998).

 \bibitem{BM} C. B\"ar and S. Moroianu, Internat. J. Math. {\bf14}, 
397 (2003).

 \bibitem{AS}  M. Abramowitz  and I. A. Stegun,   
\textit{Handbook of Mathematical Functions With Formulas,
Graphs, and Mathematical Tables} 
(U. S. Dept. of Commerce, Washington, 1964).
 
\bibitem{GR}  I. S. Gradshteyn and I. M. Ryzhik,
\textit{Table of Integrals, Series, and Products}
(Academic Press, New York, 1980).

 \bibitem{W} G. N. Watson,
 \textit{A Treatise on the Theory of Bessel Functions}, 2nd ed.
 (Cambridge U. Press, Cambridge, 1966).




\bibitem{OCF} T. A. Osborn, R. A. Corns, and Y. Fujiwara,
 J. Math. Phys. {\bf26}, 453 (1985).

 


\end{thebibliography}
\end{document}